\documentclass[journal,twocolumn]{IEEEtran}
\usepackage{graphicx}
\usepackage{epstopdf}
\usepackage{diagbox}
\usepackage{amssymb}
\usepackage{amsfonts}
\usepackage{amsmath}
\usepackage{algorithm}
\usepackage{algorithmic}
\usepackage{subeqnarray}
\usepackage{cases}
\usepackage{bm}
\usepackage{color,xcolor}

\usepackage{graphicx,subfigure,amsmath,amssymb,cite}
\usepackage{float}
\ifCLASSINFOpdf
\else
\fi
\begin{document}

\title{Two Rapid Power Iterative DOA Estimators for UAV Emitter Using Massive/Ultra-massive Receive Array}

\author{Yiwen Chen, Feng Shu, \emph{Member, IEEE}, Qijuan Jie, Xichao Zhan, Xuehui Wang, Zhongwen Sun, \\
Shihao Yan, Wenlong Cai, Peng Zhang and Peng Chen

\thanks{This work was supported in part by the National Natural Science Foundation of China (Nos.U22A2002, and 62071234), the Major Science and Technology plan of Hainan Province under Grant ZDKJ2021022, and the Scientific Research Fund Project of Hainan University under Grant KYQD(ZR)-21008.(Corresponding authors: Feng Shu).}
\thanks{Yiwen Chen, Feng Shu, Qijuan Jie, Xichao Zhan, Xuehui Wang, Zhongwen Sun, Peng Zhang, Peng Chen are with the School of Information and Communication Engineering, Hainan Unversity, Haikou, 570228, China(e-mail: shufeng0101@163.com). }
\thanks{Shihao. Yan is with the School of Science and Security Research Institute, Edith Cowan University, Perth, WA 6027, Australia (e-mail: s.yan@ecu.edu.au).}
\thanks{Wenlong  Cai is  with the National Key Laboratory of Science and Technology on Aerospace Intelligence Control, Beijing Aerospace Automatic Control Institute, Beijing 100854, China. (Email: {caiwenlon}@buaa.edu.cn).}


}

\maketitle

\begin{abstract}
To provide rapid direction finding (DF) for unmanned aerial vehicle (UAV) emitter in future wireless networks, a low-complexity direction of arrival (DOA) estimation architecture for massive multiple input multiple output (MIMO) receiver arrays is constructed. In this paper, we propose two strategies to address the extremely high complexity caused by eigenvalue decomposition of the received signal covariance matrix. Firstly, a rapid power-iterative rotational invariance (RPI-RI) method is proposed, which adopts the signal subspace generated by power iteration to gets the final direction estimation through rotational invariance between subarrays. RPI-RI makes a significant complexity reduction at the cost of a substantial performance loss. In order to further reduce the complexity and provide a good directional measurement result, a rapid power-iterative Polynomial rooting (RPI-PR) method is proposed, which utilizes the noise subspace combined with polynomial solution method to get the optimal direction estimation. In addition, the influence of initial vector selection on convergence in the power iteration is analyzed, especially when the initial vector is orthogonal to the incident wave. Simulation results show that the two proposed methods outperform the conventional DOA estimation methods in terms of computational complexity. In particular, the RPI-PR method achieves more than two orders of magnitude lower complexity than conventional methods and achieves performance close to CRLB. Moreover, it is verified that the initial vector and the relative error have a significant impact on the performance of the computational complexity.
\end{abstract}
\section{Introduction}
Unmanned aerial vehicle (UAV) plays an important role in wireless communication systems, providing stable and effective wireless connection in areas without extensive communication infrastructure coverage\cite{2015handbook,2015survey,2016survey,2021Enhanced}. Due to its economical and flexible arrangement, UAV is widely used in emergency rescue, traffic control, etc\cite{2015uav,2008airborne}. For example, UAV can provide emergency communication connection to ground rescue equipment when the mountain fire occurs. Compared with traditional ground-based communications, low-altitude UAVs have shorter line-of-sight (LoS) paths, which can effectively avoid interference and achieve better communication performance. However, its high mobility leads to rapid changes in channel state information (CSI), and ground equipment needs fast and accurate estimation of CSI to achieve high quality communication\cite{2019tutorial}.

Direction of arrival (DOA) is the key information for channel estimation\cite{Tuncer2009Classical,2018On}, which combined with multiple input multiple output (MIMO) technology\cite{2014An,2018Un,2019Ha} can achieve secure and energy-efficient UAV information transmission by providing highly accurate desired signal direction for directional modulation, beamforming, alignment and tracking\cite{shu2018secure,shu2021enhanced}. In \cite{wen20223}, the authors proposed a novel 3-D framework for UAV localization and the key of it is to measure the angle of arrival information.

However, due to the fact that the number of antennas tends to be massive in MIMO system\cite{2017massive,2019massive}, the computational complexity and circuit cost are too high for commercial applications. A DOA-aided channel estimation method was proposed in \cite{2018AN} for a hybrid millimeter-wave MIMO system based on a uniform planar array at the base stations, and the theoretical bounds of the mean squared errors (MSEs) and the Cramer-Rao lower bounds (CRLBs) of the joint DOA and channel gain estimation are derived. The simulation results show that the performances of the proposed methods are close to the theoretical MSEs’ analysis, while the theoretical MSE is close to the CRLB. Therefore, hybrid analog and digital (HAD) beamforming structures using parametric method to estimate DOA have emerged, which can achieve a good balance between beamforming computation, circuit cost, and complexity, using parametric method to estimate DOA.

Large antenna arrays using HAD architectures can provide large apertures with low cost and hardware complexity, resulting in enhanced DOA estimation and reduced power consumption. The DOA estimation and power consumption tradeoff problem for large antenna arrays with HAD structures was presented in \cite{2022DI}, where the fully-connected, sub-connected (SC), and switches-based (SE) hybrid architectures was formulated into a unified expression, with the compression matrix in a time-varying form. Based on this model, the authors derived dynamic maximum likelihood (D-ML) estimators for HAD and conventional all-digital (FD) structures, and closed expressions for CRB to evaluate the performance limits of D-ML estimators for different HAD structures.

Authors in \cite{shuDOA} investigated the DOA estimation using HAD structure in the receiver part and proposed two phase alignment (PA) methods: HAD-PA and hybrid digital and analog PA (HDA-PA). Meanwhile, for this hybrid structure, a fast Root-MUSIC-HDAPA method was proposed to achieve an approximate analytical solution and reduce the computational complexity. In \cite{2018RO}, a new design of analog phase shifts was proposed to tackle the phase ambiguities. This enables the cross-correlations to be deterministically calibrated and constructively combined for the noise-tolerant estimation of the propagation phase offset between adjacent subarrays. It is obvious from the simulation that the estimation accuracy of the method can be significantly improved by several orders of magnitude and asymptotically approaches the MSELB. For the DF ambiguity problem caused by HAD MIMO, a fast ambiguity phase elimination method was proposed in \cite{2020SBH}, which uses only two data blocks to achieve DOA estimation. In \cite{2019One}, the DOA estimation problem in the case of 1-bit ADC was considered. It demonstrated that the MUSIC method could be directly applied in the case without additional preprocessing, while the system performance degradation was analyzed. Then, the DOA estimation performance of the low-resolution ADC structure was investigated in \cite{2022Impact}.

A generalized sparse Bayesian learning (Gr-SBL) method was considered in \cite{2018AG} to solve the DOA estimation problem from one-bit quantized measurements in both single and multi snapshot scenarios. By formulating the one-bit DOA estimation in single-fast-tempo is transformed into a generalized linear model inference problem and solved by applying the recently proposed Gr-SBL method. Then, Gr-SBL is extended to multi-fast-tempo scenarios by decoupling the multi-fast-tempo single-bit DOA estimation problem into a series of single-fast-tempo subproblems. Simulation results demonstrate the effectiveness of the Gr-SBL method. A DOA estimation method that is suitable for Non-circular signals with a single snapshot was proposed in \cite{2021DOA}. By utilizing the waveform characteristics of the NC signal, the proposed algorithm can enlarge the virtual array aperture that is twice the length of the physical array, and resultantly enhance DOA estimation accuracy. Finally, the numerical simulation results are provided to demonstrate the effectiveness and superiority of the proposed method.

In order to avoid the high-complexity operation of eigenvalue decomposition (EVD) in DOA estimation, deep learning network (DNN) has been applied to DOA estimation in recent years. A DNN-based DOA and channel estimation schemes was proposed in \cite{2018Deep}, which achieved better performance. In \cite{2019Low} the authors introduced a low-complexity DNN-based method to a hybrid massive MIMO system with uniform circular array at the base station, which had similar or even better performance compared to the traditional ML method with lower complexity. An ESPRIT-based HAD method was proposed in \cite{2020Machine}, which considered a machine learning framework to improve the estimation accuracy.


Aiming at rapidly estimating and tracking the main subspaces and major components of a vector sequence in \cite{2000Fast}, a projection approximation and subspace tracking (PAST) method was proposed. Furthermore, the proposed PAST method in \cite{2000Fast} was improved in \cite{2005Fast}. It proved that the improved PAST method was better in both subspace estimation and computational complexity. In \cite{2016A}, an improved power iteration (PI) method for modal analysis was proposed. The simulation results showed that the method significantly reduced the number of unnecessary iterations with a faster computational speed. An iterative method was also proposed in \cite{2013Power}, and good results were obtained.

Inspired by the idea of radar target detection, we considered a new SVD-based passive target detection model in \cite{J-MIMO}, which achieved better detection performance. While the complexity of massive MIMO based on covariance matrix decomposition method was extremely high. For example, when the number of antennas are closed to 10000, and the computational complexity was tera(T) FLOPs. Therefore, how to significantly reduce the computational complexity of direction finding for UAV emitters is an extremely challenging problem, which is the key to its future applications. Therefore, in this paper, we have proposed a rapid convergent power iteration (RPI) structure to achieve high performance with low complexity. The main contributions in this paper are summarized as follows:

\begin{enumerate}
\item To significantly reduce the computational complexity of UAV direction finding, two PI-based DOA estimation methods are respectively proposed, which are called RPI-RI and RPI-PR. Here, the sampling covariance of the received signal vector is first computed. An initial vector subjected to power iteration is determined, which replaces the traditional EVD. Then the final DOA estimation is given by the corresponding signal and noise subspaces. The simulation results show that RPI-RI and RPI-PR can achieve better DOA accuracy and lower computational complexity than conventional algorithms, especially the RPI-PR can dramatically reduce the complexity while maintaining performance close to CRLB.
\item To reduce the number of unnecessary iterations and get a faster computation speed, the optional initial vectors are selected which can converge to the desired results. In each iteration, it must be ensured that the initial vector is not orthogonal to the incident wave, and it is better to keep them away from orthogonality. Through computational analysis, we selected different initial vector values that satisfy the conditions, and analyzed the performance of iterative convergence. Moreover, the computation result often has a great relationship with the relative error. When a good initial vector and relative error are determined, the results with fast convergence and less iterations can be achieved.
\end{enumerate}
The remainder of this paper is organized as follows. Section II describes the structure of the rapid power-iterative estimator for massive/ultra-massive MIMO receiver. In Section III, two low-complexity estimators are proposed and their performance are also analyzed. The simulation results are presented in Section IV. Finally, we draw conclusions in Section V.

\emph{Notations:} Throughout the paper, $\mathbf{x}$ and $\mathbf{X}$ in bold typeface are used to represent vectors and matrices, respectively, while scalars are presented in normal typeface, such as $x$. Signs $(\cdot)^H$ and $|\cdot|$ represent conjugate transpose and modulus, respectively. $\mathbf{I}_N$ denotes the $N\times N$ identity matrix. Furthermore, $\mathbb{E}[\cdot]$ represents the expectation operator, and $\mathbf{x}\sim \mathcal{CN}(\mathbf{m},\mathbf{R})$ denotes a circularly symmetric complex Gaussian stochastic vector with mean vector $\mathbf{m}$ and covariance matrix $\mathbf{R}$. $\hat{x}$ represents the estimated value of $x$.

\section{system model}
\begin{figure}[h]
\centering
\includegraphics[width=0.4\textwidth]{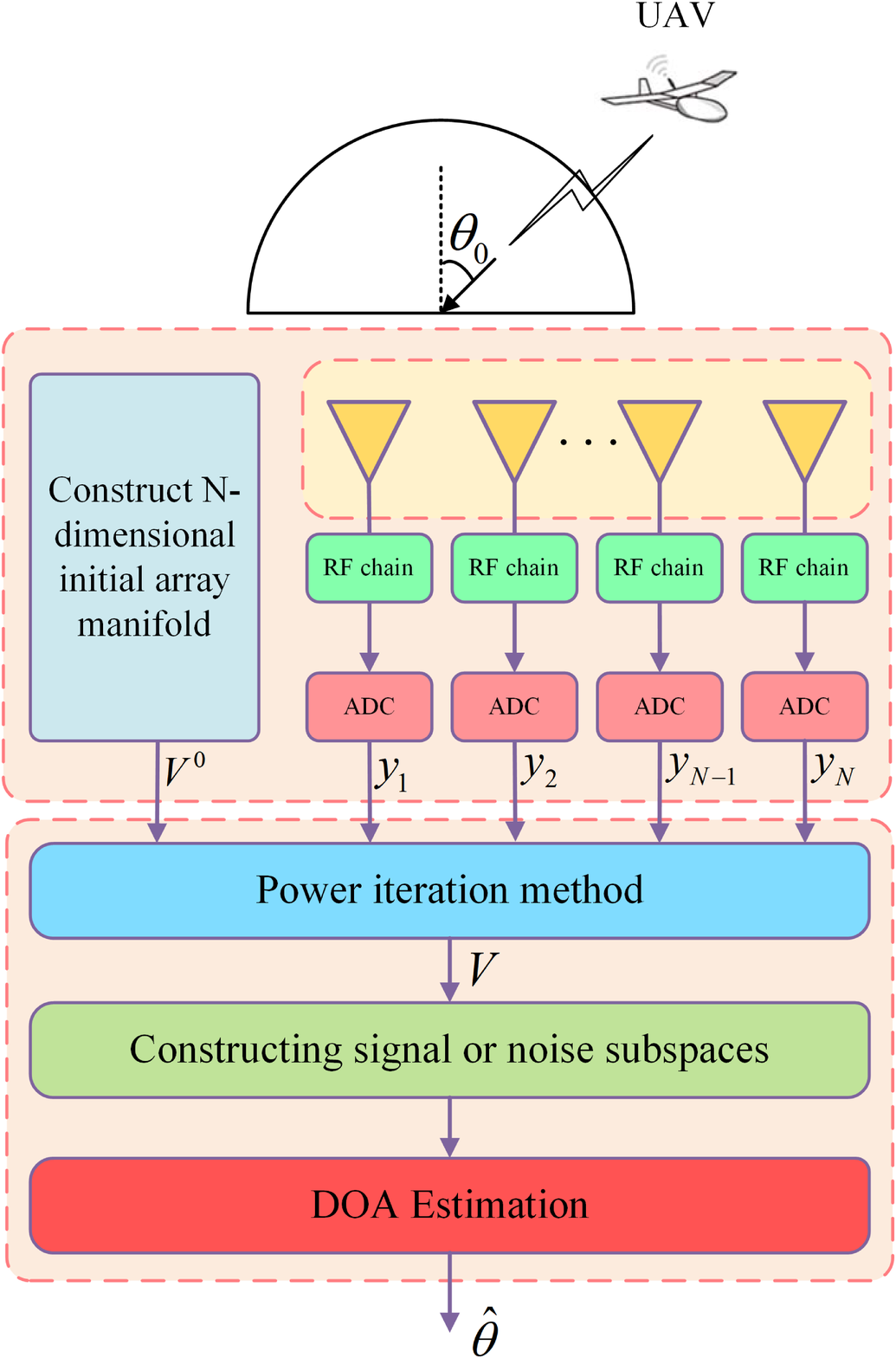}\\
\caption{Proposed a low-complexity RPI structure.}\label{sys_model.eps}
\end{figure}
To rapidly achieve the direction finding of the UAV, Fig.~\ref{sys_model.eps} sketches a low-complexity massive MIMO RPI receiver structure for UAV direction estimation. In this structure, the PI method is considered for apply in a uniformly-spaced linear array (ULA) with $N$-antennas, and an appropriate $N$-dimensional initial array manifold is constructed as the input vector for it, which can effectively reduce the number of iterations. Then, the receive covariance matrix of all antennas is exploited as the object matrix to generate an eigenvector $V$ corresponding to the largest eigenvalue. Through constructing signal or noise subspaces, the optimal DOA estimation can be derived in different methods like polynomial rooting\cite{1983Improving} and rotational invariance\cite{1986ESPRIT}.

In the proposed framework, it is assumed that the narrow band signals $s(t)e^{j2\pi f_ct}$ transmitted by UAV emitter will arrive at the array, where $s(t)$ is the baseband signal and $f_{c}$ is the carrier frequency. Then the antennas will capture the signal with different time delays depended on the DOAs. Therefore, the propagation delay of the $n$th antenna element is expressed as
\begin{align}\label{tao(n)}
\tau_{n}=\tau_{0}-(d_{n}/c){\sin}\theta_0,~~n=1,2,...,N,
\end{align}
where $\tau_{0}$ is the propagation delay from the UAV to a reference point on the receive array, $\theta_0$ is the direction of the UAV relative to the line perpendicular to the array. $d_{n}$ is the distance from the nth array element to the reference point, and $c$ is the speed of light. Without loss of generality, we can assume that $\tau_{0}=0$. Thus, $\tau_{n}=-(d_{n}/c) {\sin}\theta_0.$ The receive signal vector at array can expressed as
\begin{align}\label{y(t)}
\mathbf{y}(t)=\mathbf{a}(\theta_{0})s(t)+\mathbf{v}(t),
\end{align}
where $\mathbf{v}(t)\sim\mathcal{CN}(0,\sigma_v^2\mathbf{I}_N)$ is the additive white Gaussian noise (AWGN) vector, and $\mathbf{a}(\theta_{0})$ only composed by the phase difference of all antennas is called array manifold, defined by
\begin{align}\label{a(theta)}
\mathbf{a}(\theta_{0})=[e^{j2\pi d_{1}{sin}\theta_0/\lambda}, e^{j2\pi d_{2}{sin} \theta_0/\lambda}, \cdots, e^{j2\pi d_{N}{sin}\theta_0/\lambda}]^{T},
\end{align}

In practice, although the ideal covariance matrix can not obtained directly, its estimated value is given by
\begin{equation}\label{Ry}
\mathbf{R}_y=\frac{1}{K}\sum\limits_{n=1}^{K}\mathbf{y}(n)\mathbf{y}^{H}(n),
\end{equation}
where $K$ is the number of sampling points.

Make the eigenvalue decomposition of (\ref{Ry})
\begin{align}\label{R'}
\mathbf{R}_y=\sum\limits_{i=1}^{N}\lambda_{i}\mathbf{e}_{i}\mathbf{e}_{i}^{H}=\mathbf{U_S\Sigma_SU_S^ {H}+U_V\Sigma_VU_V^ {H}}.
\end{align}

The CRLB is the minimum variance of DOA estimation errors. It can provide a useful characterization of the achievable accuracy of the systems. According to \cite{Tuncer2009Classical}, for an $N$-element linear array we have
\begin{align}\label{CRB}
CRB=\frac{\lambda^2}{8{\pi}^2KSNR{cos}^2\theta\overline{d^2}}
\end{align}
where
\begin{align}\label{d2}
\overline{d^2}=\sum\limits_{m=1}^{N}{d_m^2}
\end{align}
and SNR is the signal-to-noise ratio of the signal received at each antenna.

\section{Proposed two rapid Power-iterative estimator for Massive/Ultra-massive MIMO Receiver}
%
%
The subspace-based methods are widely used for direction finding of UAV emitters. But their eigenvalue decomposition brings a horrible complexity when the antenna tends to large-scale. Therefore, two low-complexity estimators are proposed in this section based on the RPI structure, the selection of initial vector and computational complexity are followed analyzed.

\subsection{Proposed RPI-RI estimator}
It is assume that there exist two subarrays with both $N-1$ antennas and overlapped with each other\cite{1986ESPRIT}. The first subarray consists of the first $N-1$ antennas of all antennas, and the second subarray consists of the last $N-1$ antennas of all antennas. Since the structures of the two subarrays are identical, the outputs of the two subarrays have only one phase difference $\phi$.

The following assumes that the received data of the first subarray is $X_1$, and the received data of the second subarray is $X_2$, Combine two subarrays, the $2N-2\times 2N-2$ received data can expressed as
\begin{align}\label{XX}
\overline{\mathbf{X}}(t)=
\begin{bmatrix}
\mathbf{X}_1(t) \\
\mathbf{X}_2(t) \\
\end{bmatrix}
=
\begin{bmatrix}
\mathbf{A} \\
\mathbf{A}\bm\Phi \\
\end{bmatrix}
\mathbf{S}+\mathbf{V}=\mathbf{\overline{A}S}+\mathbf{V},
\end{align}
where $\mathbf{\Phi}=e^{j\phi}$ is a rotational invariance matrix, which contains the direction of arrival information of the incoming wave signal, and $\phi=2\pi d\sin{\theta}/\lambda$.

The covariance matrix $\mathbf{R}$ can given by
\begin{equation}\label{R}
\mathbf{R}=\frac{1}{K}\sum\limits_{n=1}^{K}\overline{\mathbf{X}}(n)\overline{\mathbf{X}}^{H}(n),
\end{equation}

Let us define $M=2N-2$ and assume that the covariance matrix $\mathbf{R}$ has $M$ eigenvalues $\lambda_{1}, \lambda_{2},...,\lambda_{M}$ with an associated collection of linearly independent eigenvectors $\{\mathbf{u_{1}}, \mathbf{u_{2}},...,\mathbf{u_{M}}\}$. Moreover, we assume that $\mathbf{R}$ has precisely one eigenvalue $\lambda_{1}$, which is the largest in magnitude, i.e.,
\begin{align}\label{all lambda}
|\lambda_{1}|>|\lambda_{2}|>|\lambda_{3}|>\cdot\cdot\cdot>|\lambda_{M}|\geq 0,
\end{align}

There exists a random vector $\mathbf{v}^{0}\in\mathbb{R}^{n}$ that satisfies
\begin{align}\label{v}
\mathbf{v}^{0}=\sum\limits_{k=1}^{N}\alpha_{k}\mathbf{u_{k}},
\end{align}
where $\alpha_{1},\alpha_{2},..., \alpha_{M}$ $(\alpha_{1}\neq 0)$ are scalars.

Taking the vector $\mathbf{v}^{0}$ as the initial vector of the RPI-RI method, and multiplying both sides of this equation by $\mathbf{R}$, ${\mathbf{R}}^{2}$, ..., ${\mathbf{R}}^{n}, ...$  gives
\begin{equation}\label{PI}
  \left\{
             \begin{array}{ll}
             \mathbf{v}^{1}=\mathbf{R}\mathbf{v}^{0}=\mathbf{R}\sum\limits_{k=1}^{M}\alpha_{k}\mathbf{u_{k}}=\sum\limits_{k=1}^{M}\alpha_{k}{\mathbf{R}}_{k}\mathbf{u_{k}}=\sum\limits_{k=1}^{M}\alpha_{k}\lambda_{k}\mathbf{u_{k}}, \\
             \mathbf{v}^{2}=\mathbf{R}\mathbf{v}^{1}={\mathbf{R}}^{2}\mathbf{v}^{0}={\mathbf{R}}^{2}\sum\limits_{k=1}^{M}\alpha_{k}\mathbf{u_{k}}=\sum\limits_{k=1}^{M}\alpha_{k}{\lambda}^{2}_{k}\mathbf{u_{k}}, \\
             \cdot\cdot\cdot \\
             \mathbf{v}^{n}=\mathbf{R}\mathbf{v}^{n-1}={\mathbf{R}}^{n}\mathbf{v}^{0}={\mathbf{R}}^{n}\sum\limits_{k=1}^{M}\alpha_{k}\mathbf{u_{k}}=\sum\limits_{k=1}^{M}\alpha_{k}{\lambda}^{n}_{k}\mathbf{u_{k}}. \\
             \end{array}
           \right.
\end{equation}

The $\mathbf{v}^{n}$ can be rewritten as
\begin{align}\label{v_{n}}
\mathbf{v}^{n}&={\lambda}^{n}_{1}(\alpha_{1}\mathbf{u_{1}}+\sum\limits_{k=2}^{M}\alpha_{k}({\frac{{\lambda}_{k}}{{\lambda}_{1}}})^{n}\mathbf{u_{k}})\nonumber\\
&={\lambda}^{n}_{1}(\alpha_{1}\mathbf{u_{1}}+\bm{\varepsilon}_{n}),
\end{align}
where
\begin{align}\label{varepsilon_{n}}
\bm{\varepsilon}_{n}=\sum\limits_{k=2}^{M}\alpha_{k}({\frac{{\lambda}_{k}}{{\lambda}_{1}}})^{n}\mathbf{u_{k}},
\end{align}

Since $|\lambda_{1}|>|\lambda_{k}|$ for all $k=2, 3, ..., M$, we have
\begin{align}
\underset{n\rightarrow\infty}{\lim}\bm{\varepsilon}_{n}=\underset{ n\rightarrow\infty}{\lim}\sum\limits_{k=2}^{M}\alpha_{k}({\frac{{\lambda}_{k}}{{\lambda}_{1}}})^{n}=0,
\end{align}
therefore,
\begin{align}
\underset{n\rightarrow\infty}{\lim}\mathbf{v}^{n}=\underset{n\rightarrow\infty}{\lim}{\lambda}^{n}_{1}\alpha_{1}\mathbf{u_{1}},
\end{align}

The eigenvector corresponding to the main eigenvalue $\lambda_{1}$ is
\begin{align}\label{SS}
\mathbf{u}_{\lambda_1}=\underset{n\rightarrow\infty}{\lim}\frac{\mathbf{v}^{n}}{{\lambda}^{n}_{1}}=\underset{n\rightarrow\infty}{\lim}\frac{{\lambda}^{n}_{1}(\alpha_{1}\mathbf{u_{1}}+\bm{\varepsilon}_{n})}{{\lambda}^{n}_{1}}=\alpha_{1}\mathbf{u_{1}},
\end{align}
where $\lambda_{1}$ is expressed by the limit of the ratio of the $i$th component of vector $\mathbf{v}^{n+1}$ to the $i$th component of vector $\mathbf{v}^{n}$, we have
\begin{align}
&\underset{n\rightarrow\infty}{\lim}\frac{(\mathbf{v}^{n+1})_i}{(\mathbf{v}^{n})_i}=\underset{n\rightarrow\infty}{\lim}\frac{{\lambda}^{n+1}_{1}(\alpha_{1}\mathbf{u_{1}}+\bm{\varepsilon}_{n+1})_i}{{\lambda}^{n}_{1}(\alpha_{1}\mathbf{u_{1}}+\bm{\varepsilon}_{n})_i}\nonumber\\
&=\underset{n\rightarrow\infty}{\lim}\frac{\lambda_{1}[\alpha_{1}(\mathbf{u_{1}})_i+(\bm{\varepsilon}_{n+1})_i]}{\alpha_{1}(\mathbf{u_{1}})_i+(\bm{\varepsilon}_{n})_i}=\lambda_{1},
\end{align}

The estimated value of $\lambda_1$ is
\begin{align}
\hat{\lambda}_{1}=\frac{(\mathbf{v}^{n+1})_i}{(\mathbf{v}^{n})_i}.
\end{align}

In cases for which the power method generates a good approximation of a dominant eigenvector $\mathbf{v}_{\lambda_1}$, the Rayleigh quotient \cite{2004Matrix} provides a correspondingly good approximation of the dominant eigenvalue $\lambda_{1}$
\begin{align}
\lambda_{1}=R(\mathbf R,\mathbf{v}_{\lambda_1})=\frac{{\mathbf{v}_{\lambda_1}}^T\mathbf R\mathbf{v}_{\lambda_1}}{{\mathbf{v}_{\lambda_1}}^T\mathbf{v}_{\lambda_1}}
\end{align}

The key problem solved by the ESPRIT algorithm is the proper use of the translation-invariant property of the linear array, so that the eigenvalues of the rotation-invariant matrix can be found to estimate the signal incidence angle.

Based on (\ref{PI}) and (\ref{SS}), we perform PI method on the covariance matrix $\mathbf{R}$ and get the signal subspace $\mathbf{u}_{\lambda_{1}}$. Since the shift invariance of the array implies that $\mathbf{u}_{\lambda_{1}}$ can be decomposed as
\begin{align}\label{}
\mathbf{u}_{\lambda_{1}}=\left[\begin{array}{l}
\mathbf{u}_{\lambda_{11}} \\
\mathbf{u}_{\lambda_{12}}
\end{array}\right]
\end{align}
where the two parts $\mathbf{u}_{\lambda_{11}}$ and $\mathbf{u}_{\lambda_{12}}$ corresponding to the signal subspaces of the subarrays $\mathbf{X}_1(t)$ and $\mathbf{X}_2(t)$. According to the ESPRIT algorithm \cite{1986ESPRIT}, the similar matrice of $\mathbf{\Phi}$ can be expressed as
\begin{align}\label{Psi}
\mathbf{\Psi}=(\mathbf{u}_{\lambda_{11}})^{\dag}\mathbf{u}_{\lambda_{12}}=({\mathbf{u}^{H}_{\lambda_{11}}}\mathbf{u}_{\lambda_{11}})^{-1}{\mathbf{u}^{H}_{\lambda_{11}}}\mathbf{u}_{\lambda_{12}},
\end{align}

Therefore, the corresponding eigenvalues of $\mathbf{\Phi}$, i.e. the diagonal elements, can be given by performing eigenvalue decomposition on $\mathbf{\Psi}$. The final DOA estimation is followed calculated by
\begin{align}\label{theta}
\hat{\theta}=\arcsin(\frac{\phi\lambda}{2\pi d}),
\end{align}
where $\phi$ is the eigenvalues of $\mathbf{\Phi}$. The specific algorithm steps of the proposed RPI-RI estimator is described in Algorithm 1 as follows
\begin{table}[htp]\normalsize
\renewcommand{\arraystretch}{1.2}
\centering
\begin{tabular}{p{240pt}}
\hline
$\bf{Algorithm~1}$ ~ Proposed RPI-RI estimator\\
\hline
$\bf{1}$: Input subarray 1 and subarray 2 to receive $\mathbf{X}_1$ and $\mathbf{X}_2$, forming the receive data model $\mathbf{\overline{X}}$ based on (\ref{XX});\\
$\bf{2}$: Calculate the covariance matrix $\mathbf{R}$ and use it as the object matrix of the power iteration to get the signal subspace corresponding to the two subarrays, and further find the similar matrix $\mathbf{\Psi}$ of $\mathbf{\Phi}$; \\
$\bf{3}$: The corresponding eigenvalues are given by EVD on $\mathbf{\Phi}$ and the final DOA estimation $\hat{\theta}$ is calculated based on (\ref{theta}). \\
\hline
\end{tabular}
\end{table}
\subsection{Proposed RPI-PR estimator}
The RPI-RI estimator can significantly reduce the computational complexity by using the power iteration and the rotational invariance of the signal subspace between subarrays, but it is difficult to achieve the desired performance and the complexity can be further optimized.

In order to further reduce the computational complexity and achieve better performance, RPI-PR estimator is proposed in this subsection. From (\ref{SS}), the signal subspace can given by power iterating over $\mathbf{R}_y$ in (\ref{Ry}). Furthermore, we construct the noise subspace
\begin{align}\label{V}
\mathbf{U}_V=\mathbf{I}-\mathbf{u}_{\lambda_1}(\mathbf{u}_{\lambda_1}^{H}\mathbf{u}_{\lambda_1})^{-1}\mathbf{u}_{\lambda_1}^{H},
\end{align}
where $\mathbf{I}$ is $N\times N$ unit matrix. The spatial spectral function can defined as
\begin{align}\label{S}
S(\theta)=\frac{1}{\|\mathbf{a}^H(\theta)\mathbf{U}_V\|^2}
\end{align}
which spectral peak corresponds to the desire DOA estimation. Furthermore, let us define $z=e^{2\pi/\lambda dsin\theta}$, the polynomial equation can be expressed as
\begin{align}\label{f}
f(\theta) & =S^{-1}(\theta)=\mathbf{a}^H(\theta)\mathbf{U}_V\mathbf{U}^H_V\mathbf{a}(\theta)\\\nonumber
&=\mathbf{a}^T(\frac{1}{z})\mathbf{U}_V\mathbf{U}^H_V\mathbf{a}(z)\triangleq f(z)
\end{align}

 The above polynomial equation (\ref{f}) has $2N-2$ roots, i.e $z_i,i=1,\cdots,2N-2$, which implies the existence of multiple emitter directions as follows
\begin{align} \label{}
\hat{\Theta}_{R M}=\left\{\hat{\theta}_i, i \in\{1,2, \cdots, 2 N-2\}\right\},
\end{align}
where
\begin{align} \label{theta}
\hat{\theta}_i=\arcsin \left(\frac{\lambda \arg z_i}{2 \pi d}\right)
\end{align}

The angle corresponding to the root inside the unit circle and closest to it is chosen as the final DOA estimation $\hat{\theta}$.

And the basic steps of the proposed RPI-PR method can be summarized in Algorithm 2 as
\begin{table}[htp]\normalsize
\renewcommand{\arraystretch}{1.2}
\centering
\begin{tabular}{p{240pt}}
\hline
$\bf{Algorithm~2}$ ~ Proposed RPI-PR method\\
\hline
$\bf{1}$: Input array receive signal y(t);\\
$\bf{2}$: Based on (\ref{Ry}), calculate the covariance matrix $\mathbf{R}_y$; \\
$\bf{3}$: Based on (\ref{V}),to estimate the noise subspace $\mathbf{U}_V$ and use it to construct the spatial spectrum $S(\theta)$   \\
$\bf{4}$: Multiple roots are given by polynomial method for the spatial spectral function, the angle corresponding to the root inside the unit circle and closest to it is chosen as the final DOA estimation $\hat{\theta}$.  \\
\hline
\end{tabular}
\end{table}
\subsection{Selecting initial vector and relative error}
The initial vector $\mathbf{v}^{0}$ has a direct effect on the speed of convergence and determines the number of iterations. A randomly generated vector is usually selected as the iterative initial vector, but not all the optional initial vectors can converge to get the desired results. Since orthogonality may cause the iterative process fails to converge. Therefore, it is necessary to ensure that the initial vector is not orthogonal to the incident wave in each iteration.

From (\ref{v}) and (\ref{v_{n}}), the $\mathbf{v}^{0}$ is formed as
\begin{align}\label{v'}
(\mathbf{v}^{0})^{H}=\frac{\sum\limits_{j=1}^{N}\mathbf{R}_{.j}}{\mathbf {S(R)}}
\end{align}
where $\mathbf {S(R)}$ is the sum of all elements in matrix $\mathbf{R}$,
\begin{align}
\mathbf {S(R)}=\sum\limits_{i=1}^{N}\sum\limits_{j=1}^{N}\mathbf{R}_{ij}
\end{align}
and
\begin{align}
\sum\limits_{j=1}^{N}\mathbf{R}_{.j}=\mathbf {1}^{H}_{N\times1}\mathbf{R}
\end{align}

Thus, we can calculate
\begin{align}
\mathbf{R}\mathbf{v}^{0}&=\mathbf{R}\frac{(\sum\limits_{j=1}^{N}\mathbf{R}_{.j})^{H}}{\mathbf {S(R)}}=\mathbf{R}(\mathbf {1}^{H}_{N\times1}\mathbf{R})^{H}/\mathbf {S(R)}\nonumber\\
&=\mathbf{R}\mathbf{R}^{H}\mathbf {1}_{N\times1}=\mathbf {1}_{N\times1}
\end{align}

Based on the above discussion, the element distribution of initial vector $\mathbf{v}^{0}$ in (\ref{v'}) is consistent with the distribution of matrix $\mathbf{R}$, while the distribution is uniform, which can keep them far away from orthogonality and speed up the convergence of iteration.

Therefore, let us define the array manifold as
\begin{align}\label{a(theta0)}
\mathbf{a}(\theta_{0})&=[1, e^{-j2\pi\frac{d}{\lambda}{sin}\theta_0}, \cdots, e^{-j2\pi(N-1)\frac{d}{\lambda}{sin}\theta_0}]^{T}\nonumber\\
&=[1, e^{-j\phi}, e^{-j2\phi}, \cdots, e^{-j(N-1)\phi}]^{T},
\end{align}
where
\begin{align}
\phi=2\pi\frac{d}{\lambda}{sin}\theta_0
\end{align}
and the initial vector $\mathbf{v}^{0}$ is assumed as
\begin{align}\label{v^0}
\mathbf{v}^{0}=[b_0, b_1, \cdots, b_{N-1}]^{T},
\end{align}

The orthogonality equation can be expressed as
\begin{align}\label{av}
\mathbf{a}(\theta_{0})^{T}\mathbf{v}^{0}=b_0+b_1e^{-j\phi}+...+b_{N-1}e^{-j(N-1)\phi}
\end{align}

In order to make the incident wave direction not orthogonal to the initial vector, i.e., the $\mathbf{a}(\theta_{0})^{T}\mathbf{v}^{0}\neq0$ is constant. In the following, eight special initial vectors are discussed and their convergence performance is analyzed.

I.
The initial vector is assumed as
\begin{align}\label{34}
\mathbf{v}^{0}=[1, 1, 1, \cdots, 1]^{T}.
\end{align}
where all elements are 1. Based on (\ref{av}), corresponding orthogonality equation is given by
\begin{align}\label{av1}
\mathbf{a}(\theta_{0})^{T}\mathbf{v}^{0}&=1+e^{-j\phi}+e^{-j2\phi}+...+e^{-j(N-1)\phi}\nonumber\\
&=\frac{1-(e^{-j\phi})^N}{1-e^{-j\phi}}
\end{align}
where $\phi=2\pi\frac{d}{\lambda}{sin}\theta_0$ and it can be discussed in three cases.

(1) When $1-e^{j\phi}\neq0$ and $1-(e^{-j\phi})^N=0$, $\phi$ can be calculated as
\begin{align}\label{phi1}
	\phi=2\pi\frac{d}{\lambda}{sin}\theta_0=\frac{2\pi Z}{N}
\end{align}
where $Z$ is integer. It can be seen from (\ref{phi1}) that $\mathbf{a}(\theta_{0})^{T}\mathbf{v}^{0}=0$ holds when $\theta_0=arcsin(\frac{\lambda Z}{Nd})$ and $\frac{Z}{N}$ is not integer.

(2) When $1-e^{j\phi}=0$, $\phi$ is given by
\begin{align}
	\phi=2\pi\frac{d}{\lambda}{sin}\theta_0=2\pi Z
\end{align}

Substituting $\phi=2\pi Z$ into (\ref{av1}), $\mathbf{a}(\theta_{0})^{T}\mathbf{v}^{0}\neq0$ is verified to hold.

(3) When $1-e^{j\phi}\neq0$ and $1-(e^{-j\phi})^N\neq0$, it is clear that $\mathbf{a}(\theta_{0})^{T}\mathbf{v}^{0}\neq0$ holds. Through the above discussion, initial vector I is not guaranteed non-orthogonal always true.

II. The initial vector is assumed as
\begin{align}\label{39}
\mathbf{v}^{0}=[1, -1, 1, -1, \cdots]^{T}.
\end{align}
where the odd elements is 1 and the even elements is -1. Depending on the value of $N$, two cases can be discussed.

(1) $N$ is even
\begin{align}
\mathbf{a}(\theta_{0})^{T}\mathbf{v}^{0}&=1-e^{-j\phi}+e^{-j2\phi}+...-e^{-j(N-1)\phi}\nonumber\\
&=(1+e^{-j2\phi}+e^{-j4\phi}+...+e^{-j(N-2)\phi})\nonumber\\
&~~~~-(e^{-j\phi}+e^{-j3\phi}+...+e^{-j(N-1)\phi})\nonumber\\
&=\frac{1-(e^{-j2\phi})^{\frac{N}{2}}}{1-e^{-j2\phi}}-\frac{e^{-j\phi}(1-(e^{-j2\phi})^{\frac{N}{2}})}{1-e^{-j2\phi}}\nonumber\\
&=\frac{(1-e^{-j\phi})(1-(e^{-j2\phi})^{\frac{N}{2}})}{1-e^{-j2\phi}}
\end{align}

From the above discussion, it is clear that $\mathbf{a}(\theta_{0})^{T}\mathbf{v}^{0}=0$ holds when $\phi$ satisfies the following form
\begin{equation}
  \left\{
             \begin{array}{ll}
           i.~~\phi=2\pi N,~N~is~integer\\
           ii.~~\phi=\frac{2\pi Z}{N},\frac{Z}{N}~is~not~integer.\\
             \end{array}
           \right.
\end{equation}

If $\phi$ selects these cases, it may generate incorrect DOA estimation.

(2) $N$ is odd
\begin{align}
\mathbf{a}(\theta_{0})^{T}\mathbf{v}^{0}&=1-e^{-j\phi}+e^{-j2\phi}+...+e^{-j(N-1)\phi}\nonumber\\
&=(1+e^{-j2\phi}+e^{-j4\phi}+...+e^{-j(N-1)\phi})\nonumber\\
&~~~~-(e^{-j\phi}+e^{-j3\phi}+...+e^{-j(N-2)\phi})\nonumber\\
&=\frac{1-(e^{-j2\phi})^{\frac{N+1}{2}}}{1-e^{-j2\phi}}-\frac{e^{-j\phi}(1-(e^{-j2\phi})^{\frac{N-1}{2}})}{1-e^{-j2\phi}}\nonumber\\
&=\frac{(1-e^{-j\phi})(1+e^{-jN\phi})}{1-e^{-j2\phi}}
\end{align}

It can be seen that $\mathbf{a}(\theta_{0})^{T}\mathbf{v}^{0}=0$ holds when $\phi$ satisfies the following conditions
\begin{align}
\phi=\frac{(2n+1)\pi}{N},
\end{align}
where $\frac{2n+1}{N}$ is not integer and orthogonality only can be avoided by selecting other $\phi$.

III. The initial vector is assumed as
\begin{align}\label{44}
\mathbf{v}^{0}=[1, 1, 1, 1, \cdots, -1, -1, \cdots, -1]^{T}.
\end{align}

Assuming $N$ is even. The first half of the elements of vector $\mathbf{v}^{0}$ are 1, and the rest are -1.
\begin{align}
\mathbf{a}(\theta_{0})^{T}\mathbf{v}^{0}&=1+e^{-j\phi}+...-e^{-j(N-1)\phi}\nonumber\\
&=(1+e^{-j\phi}+e^{-j2\phi}+...+e^{-j(\frac{N}{2}-1)\phi})\nonumber\\
&~~~~-(e^{-j\frac{N}{2}\phi}+...+e^{-j(N-1)\phi})\nonumber\\
&=\frac{1-(e^{-j\phi})^{\frac{N}{2}}}{1-e^{-j\phi}}-\frac{e^{-j\frac{N}{2}\phi}(1-(e^{-j\phi})^{\frac{N}{2}})}{1-e^{-j\phi}}\nonumber\\
&=\frac{(1-e^{-j\frac{N}{2}\phi})(1-(e^{-j\phi})^{\frac{N}{2}})}{1-e^{-j\phi}}
\end{align}

Therefore, $\mathbf{a}(\theta_{0})^{T}\mathbf{v}^{0}=0$ holds when $\phi$ satisfies the following cases

\begin{equation}
  \left\{
             \begin{array}{ll}
           i.~~\phi=2\pi N,~N~is~integer\\
           ii.~~\phi=\frac{4\pi Z}{N},\frac{2Z}{N}~is~not~integer.\\
             \end{array}
           \right.
\end{equation}

The above value of $\phi$ will lead to orthogonality.

IV. Let us define $v_0$ as the normalized DFT vector $\mathbf{f}$ with its elements are  given by \cite{2017A}
\begin{align}\label{48}
	\mathbf{v}^{0}=\mathbf{f}&=[e^{-j\frac{2\pi}{N}/\sqrt{N}}, e^{-j\frac{2\pi}{N}2/\sqrt{N}}, \cdots, e^{-j\frac{2\pi}{N}N/\sqrt{N}}]^{T}\nonumber\\
	&=[e^{-jA}, e^{-j2A}, \cdots, e^{-jNA}]^{T}.
\end{align}

where
\begin{align}
A=\frac{2\pi}{N}/\sqrt{N}
\end{align}

Therefore, (\ref{av}) can be calculate as
\begin{align}
\mathbf{a}(\theta_{0})^{T}\mathbf{v}^{0}&=e^{-jA}+e^{-j2A}e^{-j\phi}+...+e^{-jNA}e^{-j(N-1)\phi}\nonumber\\
&=e^{-jA}+e^{-j(2A+\phi)}+...+e^{-j(NA+(N-1)\phi)}\nonumber\\
&=\frac{e^{-jA}(1-(e^{-jA}e^{-j\phi})^N)}{1-(e^{-jA}e^{-j\phi})}
\end{align}

Solving the above equation yields
\begin{align}
&\phi=\frac{2\pi Z}{N}-A=2\pi\frac{d}{\lambda}{sin}\theta_0\nonumber\\
&{sin}\theta_0=\frac{\lambda(Z\sqrt{N}-1)}{N\sqrt{N}d}.
\end{align}

From above discussion, when ${sin}\theta_0=\frac{\lambda(Z\sqrt{N}-1)}{N\sqrt{N}d}$, where $\frac{Z}{N}$ is not integer, we have $\mathbf{a}(\theta_{0})^{T}\mathbf{v}^{0}=0$, which orthogonality may lead to estimation failure.

V.
The initial vector is assumed as
\begin{align}\label{two}
\mathbf{v}^{0}=[\cdots, 1, 0, 0, 0, \cdots, 1, \cdots]^{T}, m,n=1,\cdots,N.
\end{align}
where $m$-th element is 1, the $n$-th is 1 and the rest of the elements are 0 ($m<n$).
\begin{align}
\mathbf{a}(\theta_{0})^{T}\mathbf{v}^{0}=e^{-jm\phi}+e^{-jn\phi}=e^{-jm\phi}(1+e^{-j(n-m)\phi})
\end{align}

When $\phi=\frac{(2Z+1)\pi}{n-m}$, $\mathbf{a}(\theta_{0})^{T}\mathbf{v}^{0}=0$.

VI.
The initial vector is assumed as
\begin{align}\label{three}
\mathbf{v}^{0}=[\cdots, 1, 0, \cdots, 1, \cdots, 1, \cdots]^{T}, m,n,k=1,\cdots,N.
\end{align}
where $m,n,k$-th is 1 and the rest of the elements are 0 ($m<n<k$).
\begin{align}
\mathbf{a}(\theta_{0})^{T}\mathbf{v}^{0}&=e^{-jm\phi}+e^{-jn\phi}+e^{-jk\phi}\nonumber\\
&=e^{-jm\phi}(1+e^{-j(n-m)\phi}(1+e^{-j(k-n)\phi}))
\end{align}

When $\phi=\frac{Z\pi}{n-m}$ or $\frac{(2Z+1)\pi}{k-n}$, $\mathbf{a}(\theta_{0})^{T}\mathbf{v}^{0}=0$.

VII.
The initial vector is assumed as
\begin{align}\label{four}
&\mathbf{v}^{0}=[\cdots, 1, 0, \cdots, 1, \cdots, 1, \cdots, 1, \cdots]^{T},\\\nonumber
&m,n,k,l=1,\cdots,N.
\end{align}
where $m,n,k,l$-th is 1 and the rest of the elements are 0 ($m<n<k<l$).
\begin{align}
\mathbf{a}(\theta_{0})^{T}\mathbf{v}^{0}&=e^{-jm\phi}+e^{-jn\phi}+e^{-jk\phi}+e^{-jl\phi}\nonumber\\
&=e^{-jm\phi}.\nonumber\\
&(1+e^{-j(n-m)\phi}(1+e^{-j(k-n)\phi}(1+e^{-j(l-k)\phi})))
\end{align}

When $\phi=\frac{Z\pi}{n-m}$ or $\frac{Z\pi}{k-n}$ or $\frac{(2Z+1)\pi}{l-k}$, $\mathbf{a}(\theta_{0})^{T}\mathbf{v}^{0}=0$.

So we can get the general rule, when $b_1,b_2,b_3,\ldots,b_n$ is 1 and the rest of the elements are 0 ($b_1<b_2<\ldots<b_n$), when $\phi=\frac{Z\pi}{b_2-b_1}$ or $\frac{Z\pi}{b_3-b_2}$ or $\ldots$ $\frac{Z\pi}{b_{n-1}-b_{n-2}}$ or $\frac{(2Z+1)\pi}{b_n-b_{n-1}}$, $\mathbf{a}(\theta_{0})^{T}\mathbf{v}^{0}=0$.

VIII.
The initial vector is assumed as
\begin{align}\label{one}
\mathbf{v}^{0}=\mathbf{e}_{k}=[\cdots, 0, 0, \cdots, 1, \cdots ]^{T}, k=0,\cdots,N-1.
\end{align}
where k-th element of $\mathbf{e}_{k}$ is 1, and the rest of the elements are 0.
$\mathbf{a}(\theta_{0})^{T}\mathbf{v}^{0}\neq0$ is always true.

Among the above eight vector forms, the eighth vector form can be used as the initial vector of the power iteration due to its non-orthogonality, while the previous seven cannot guarantee the convergence of the power iteration.

In addition, the relative error $\varepsilon$ also influence the speed of convergence in the convergence process of power iteration, which can be expressed as
\begin{align}
\varepsilon=\frac{|\bm{\varepsilon}_{n}|}{|\alpha_{1}\mathbf{u_{1}}|}=\frac{|\sum\limits_{k=2}^{N}\alpha_{k}({\frac{{\lambda}_{k}}{{\lambda}_{1}}})^{n}\mathbf{u_{k}}
|}{|\alpha_{1}\mathbf{u_{1}}|}
\end{align}
and
\begin{align}
|\sum\limits_{k=2}^{N}\alpha_{k}({\frac{{\lambda}_{k}}{{\lambda}_{1}}})^{n}|\leq |\sum\limits_{k=2}^{N}\alpha_{2}({\frac{{\lambda}_{2}}{{\lambda}_{1}}})^{n}|
\end{align}

Therefore, we have
\begin{align}
&\varepsilon\leq (N-1)({\frac{|{\lambda}_{2}|}{|{\lambda}_{1}|}})^{n}\frac{\alpha_{2}}{\alpha_{1}},\nonumber\\
&\log_2{\varepsilon}\leq \log_2(N-1)+n\log_2{({\frac{|{\lambda}_{2}|}{|{\lambda}_{1}|}})}+\log_2{\frac{\alpha_{1}}{\alpha_{2}}}
\end{align}

Available from $\alpha_{1}\geq \alpha_{2}$,
\begin{align}
\log_2{\varepsilon}\leq\log_2(N-1)+n\log_2{({\frac{|{\lambda}_{2}|}{|{\lambda}_{1}|}})}
\end{align}

Since $|\lambda_{1}|>|\lambda_{2}|$,
\begin{align}
n\leq\frac{\log_2{\varepsilon}-\log_2(N-1)}{\log_2{({\frac{|{\lambda}_{2}|}{|{\lambda}_{1}|}})}}
\end{align}

When the iteration approaches the final convergence, the convergence speed will be relatively slow and stable. In practice, the usefulness of the power method depends upon the ration $\frac{|{\lambda}_{2}|}{|{\lambda}_{1}|}$, since it dictates rate of convergence.

The whole procedure is summarized in Algorithm 3.
\begin{table}[h]\normalsize
\renewcommand{\arraystretch}{1}
\centering
\begin{tabular}{p{240pt}}
\hline
$\bf{Algorithm~3}$ ~ PI method on subspace\\
\hline
$\bf{Input}$: $\bf{R}$ and $\varepsilon$ \\
$\bf{Output}$: $\lambda_1$,~$\mathbf{v}^{n}$ and n \\
$\bf{Initialization}$: choose a initial vector $\mathbf{v}^{0}$,~and n=1. \\
$\bf{repeat}$ \\
~~~1. $n$ = $n$+1.\\
~~~2. Update $\mathbf{v}^{0}$,~$\mathbf{v}^{0}$=$\mathbf{v}^{1}$;\\
~~~3. Update $\mathbf{v}^{1}$;\\
$\bf{until}$ \\
\ \ \ \ \    $\|\bf{max}(\mathbf{v}^{0})-\bf{max}(\mathbf{v}^{1})\|\le \varepsilon$.\\
$\bf{Return}$ $\lambda_1$,~$\mathbf{v}^{n}$ and n\\
\hline
\end{tabular}
\end{table}
where n is the number of iterations, $\lambda_1$ is dominant eigenvalue, and $\mathbf{v}^{n}$ is the dominant eigenvector corresponding to the $\lambda_1$.

Notice that in each iteration we compute a single matrix-vector multiplication $(O(N^2))$. We never perform matrix-matrix multiplication, which requires greater number of operations $(O(N^3))$. If the matrix R is sparse (only a small portion of the entries of A are non-zero), matrix-vector multiplication can be performed very efficiently. Therefore, the power method is practical even if $N$ is very large, such as in Google's Page Rank algorithm.
\subsection{Complexity analysis}
we make an analysis of computational complexities of the proposed two estimators with traditional ESPRIT and Root-MUSIC algorithms as a complexity benchmark.
Thus, the complexity of RPI-RI is as follows
\begin{align}
C_{RPI-RI} = O\{\beta(2N-2)^2+2N-3\}
\end{align}
FLOPs. The complexity of RPI-PR is
\begin{align}
C_{RPI-PR} = O\{(\beta+8)N^2+NK(2N+3)-11N+4\}
\end{align}
FLOPs, where $\beta$ is iteration number of the power iteration. Assuming $N$ is far larger than $K, \beta$, compared with the conventional FD estimator, the complexity of the proposed two estimators is significantly reduced as the number of antennas tends to large-scale.
\section{Simulation results and discussions}
In this section, we provide the simulation to analyze the performance of the proposed massive MIMO DOA estimators for UAV and two conventional algorithms are used as a comparison. Furthermore, we consider the effect of SNR, the number of antennas and the number of snapshots on the proposed methods in the digital ADC architecture. In each simulation figure, the number of estimation value $L$ is 2000. Without loss of generality, we assume that UAV emitter located in $\theta_0= 50^o$, the number of snapshots $K$ is 1000, antennas distance $d=\lambda/2$, and the number of antenna elements $N$ is 64 in the massive MIMO system.
\begin{align}
RMSE=\sqrt{\frac{1}{L}\sum\limits_{l=1}^{L}({\hat\theta_{l}-\theta_0})^2}
\end{align}

\begin{figure}[h]
\centering
\includegraphics[width=0.5\textwidth]{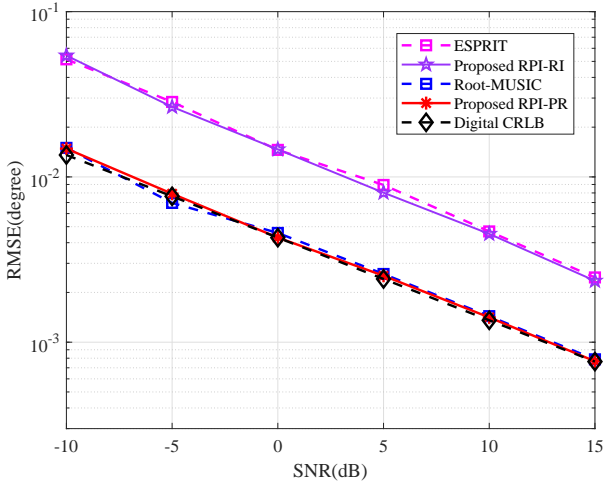}\\
\caption{RMSE over the SNR with $N=64$ and $K=1000$.}\label{SNR.eps}
\end{figure}

\begin{figure}[h]
\centering
\includegraphics[width=0.5\textwidth]{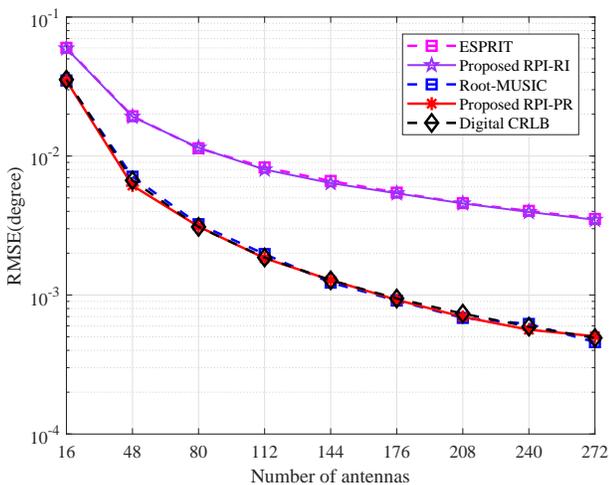}\\
\caption{RMSE over the $N$ with SNR=0dB and $K=1000$.}\label{N.eps}
\end{figure}

\begin{figure}[h]
\centering
\includegraphics[width=0.5\textwidth]{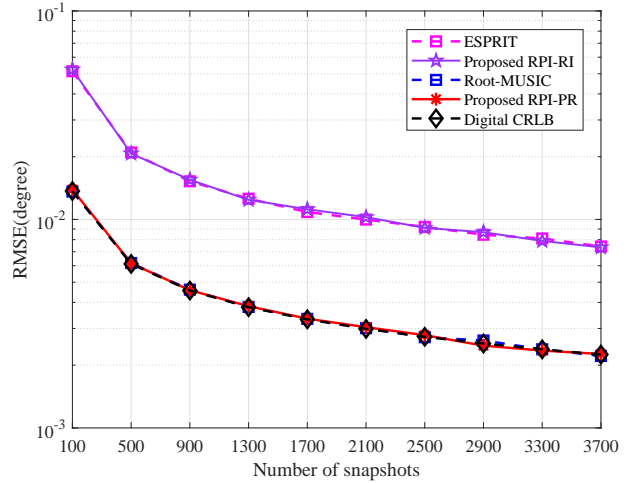}\\
\caption{RMSE over the $K$ with $N=64$ and SNR=0dB.}\label{Snapshots.eps}
\end{figure}

\begin{figure}[h]
\centering
\includegraphics[width=0.5\textwidth]{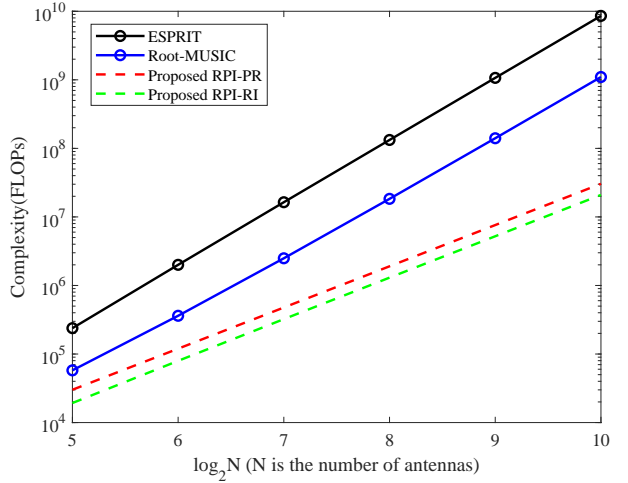}\\
\caption{Computational complexity over the number of antennas $N$ with SNR=0dB and $K=1000$}\label{complexity.eps}
\end{figure}

\begin{figure}[h]
	\centering
	\includegraphics[width=0.5\textwidth]{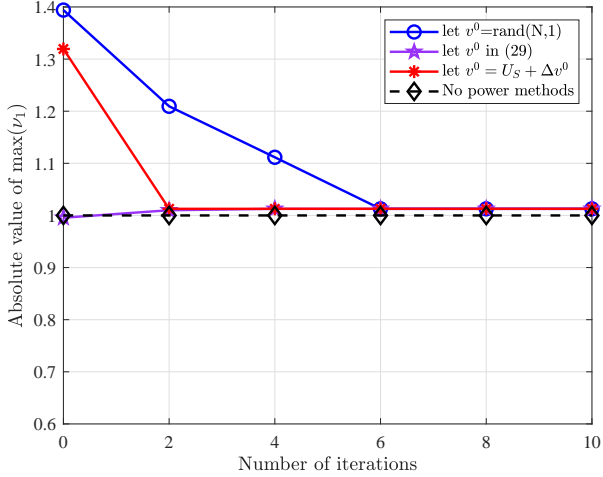}\\
	\caption{Convergence over the number of iterations $n$ with $N=64,$ $K=1000$ and SNR=0dB}\label{Iteration.eps}
\end{figure}

\begin{figure}[h]
\centering
\includegraphics[width=0.5\textwidth]{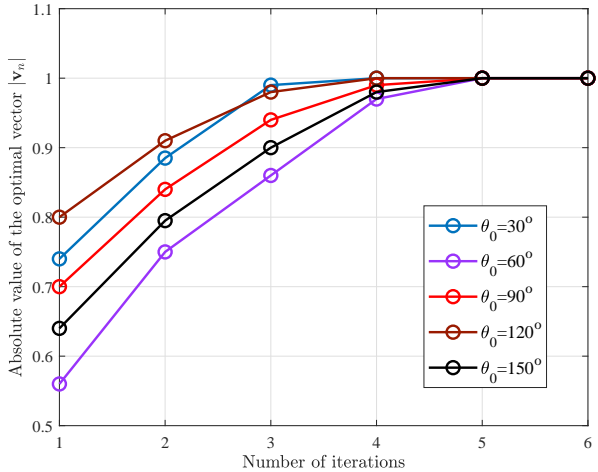}\\
\caption{Curves of the number of iterations for five different directions with $N=64$ and SNR=0dB}\label{ITERA2.eps}
\end{figure}

\begin{figure}[h]
\centering
\includegraphics[width=0.5\textwidth]{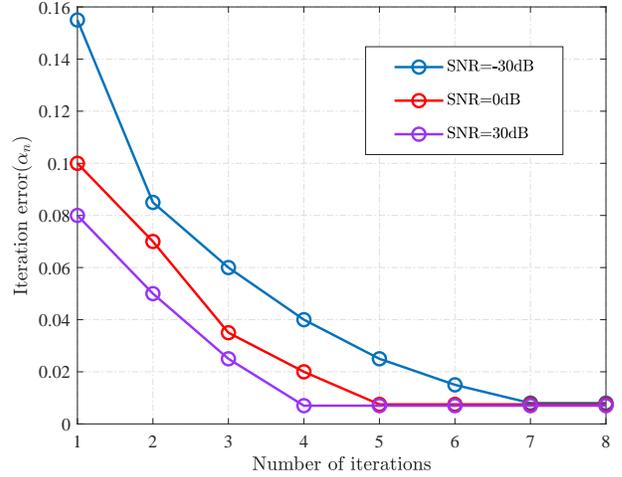}\\
\caption{Curves of the number of iterations with different SNR}\label{ITERA3.eps}
\end{figure}

Fig.~\ref{SNR.eps} plots the root mean square error (RMSE) curves using four different methods versus SNR with CRLB as the performance benchmark. Observing Fig.~\ref{SNR.eps}, it is clear that the proposed RPI-RI method can achieve a similar performance to ESPRIT but with some performance loss, while the proposed RPI-PR method can achieve the corresponding CRLB.

Fig.~\ref{N.eps} and Fig.~\ref{Snapshots.eps} present the RMSE of four different methods versus the number of antennas $N$ and snapshots $K$. Without loss of generality, we assume that $N \in[16,272]$ and $K \in [100,3600]$. From the Fig.~\ref{N.eps} and Fig.~\ref{Snapshots.eps}, it can be seen that the DOA estimators on the power iteration method achieves a performance close to the conventional algorithms for any number of antennas and snapshots scenarios, especially RPI-PR can reach the FD CRLB.

Fig.~\ref{complexity.eps} plots curves of complexity analysis versus the number of antennas $N$ with $K=50$, SNR=0dB. From this figure, it can be seen that the complexity of all methods gradually increases as the total number of antennas increases. However, the computational complexity of our proposed methods is two to three orders of magnitude lower when $N$ = 1024 compared to the conventional methods. In particular, the Proposed RPI-PR method can achieve a performance close to CRLB.

To explore the influence of the selection of the initial vector on the number of iterations and the convergence speed, Fig.~\ref{Iteration.eps} shows the relationship between the optimal eigenvector value $\mathbf{\nu^n}$ and the number of iterations $n$, given three different initial vectors $\mathbf{\nu^0}$. When the initial vector $\mathbf{\nu^0}$ is infinitely close to the signal subspace, only two iterations are needed to complete the convergence, and the result is similar to that of the initial vector selected in (\ref{v'}). In addition, when the $\mathbf{\nu^0}$ obeys a random vector distribution, the required number of iterations $n$ is more, which requires an average of eight iterations to reach the convergence of the $max(\nu^n)$.


In order to verify the convergence performance of the initial vector $\mathbf{v}^{0}=\mathbf{e}_{k}=[\cdots, 0, 0, \cdots, 1, \cdots ]^{T}$ in (\ref{one}), five incident wave directions $\theta_0=\{30^o,60^o,90^o,120 ^o,150^o\}$ are assumed in Fig.~\ref{ITERA2.eps} and the number of iterations of RPI methods is given by numerical simulation when SNR=0. As shown in Fig.~\ref{ITERA2.eps}, the proposed RPI methods requires about five iterations to converge. The $|\mathbf{v}_n|$ in the figure indicates the absolute value of the vector at the $n$th iteration.

To analyze the effect of SNR on the convergence speed of the proposed RPI methods. Fig.~\ref{ITERA3.eps} plots the iteration error $\alpha_n$ of the RPI methods versus the number of iterations, where the iteration error $\alpha_n$ represents the absolute value of the difference between the result of the $n$th iteration $C(\theta_n)$ and the result of the $(n-1)$th iteration $C(\theta_{n-1})$, i.e.$\alpha_n=|C(\theta_n)-C(\theta_{n-1})|$.  As shown in Fig.~\ref{ITERA3.eps}, assuming the number of antennas $N=1024$ and the incident wave direction $\theta_0= 50^o$. The number of iterations is 7, 5 and 4 for SNRs of -30dB, 0dB and 30dB, respectively. the number of iterations of the proposed algorithm decreases sequentially as the SNR increases.
\section{Conclusions}
To find the direction of UAV emitter rapidly and accurately, two low-complexity DOA estimators based on large-scale MIMO arrays are proposed. By determining good initial vector and relative errors, the proposed two algorithms can achieve fast convergence and lower complexity than conventional algorithms. In particular, the PI-PR method achieves more than two orders of magnitude complexity reduction and maintains performance close to CRLB. Adopting the proposed algorithms makes fast direction finding of UAV based on massive MIMO receiver feasible for future practical applications.

\ifCLASSOPTIONcaptionsoff
  \newpage
\fi

\bibliographystyle{IEEEtran}
\bibliography{reference}
\end{document}